# Monitoring Health Using IoT and Thingspeak

[1]Nnamdi, Micky Chisom,  [2]Joboson, Peter Kanene, [3]Dyaji, Charles Bala

[1,2,3] Electrical & Electronics Engineering Department, Air Force Institute of Technology, Kaduna, Nigeria
Email: nnamdimicky112@gmail.com; jobosonpk@gmail.com; dyaji@afit.edu.ng

**Abstract**

*Quality of health is greatly impacted by the quality of measurement, hence active monitoring of patient's vitals status greatly affects recovery and improves prognosis. In this paper, we proposed an IoT health based monitoring system which comprises of a wearable wristband mounted on the patient wrist to eliminate the concept of patient's being static. The wearable wristband composed of a digital temperature sensor and a pulse rate sensor can remotely monitor health status and utilizing Thingspeak result viewed via a web app or delivered on a smartphone to a specialist who is in a remote location. In order to achieve high accuracy in readings, the signal measured from the pulse rate sensor was passed through a two stage Hp-Lp circuit to eliminate DC offset and noise. In addition, a scaling factor of 10.41909 was formulated and used to convert the digital output from the pulse rate sensor into a more accurate representation of the patient heat rate readings. The results displayed by the proposed system proved the system to be well suited and reliable for healthcare monitoring.*

## 1. Introduction

The word health gets its originality from the field of healthcare and is used mainly to describe the state of a living thing. Monitor means a device used for observing, checking or to keep a continuous record of something and remote means operating or controlling at a distance. Therefore, a remote health monitoring system is a device which checks and observes the essential state of a patient's body remotely, that is at a distance. Originally, the first set of health monitoring systems found its application only in the hospitals. Also, this system was huge and had complex circuitry that required high power to operate properly.  The continuous advancement in the semi-conductor technology industry led to the development of microcontrollers and sensors that are minute, works faster, needs just low power usage and cost efficient. This remote health monitoring system tends to observe parameters like temperature, blood pressure, heart rate, electrocardiography ECG, etc. using relevant sensors and then communicate them through Bluetooth, IoT, GPRS etc. The types of remote health monitoring system depend basically on the type of parameters being measured or observed e.g.

- Temperature based monitoring system
- Temperature and heart rate-based monitoring system
- Blood pressure-based monitoring system etc. (TAN et al, 2012)

IoT came to light in the early 2000s during a presentation by Kevin Ashton for PROCTOR and GAMBLE, he suggested a system whereby objects in the physical realm with sensors embedded in it could be connected through the internet  (Ashton, 2009). Currently, IoT is a known term used to describe cases in which connectivity and computing of internet capability gets to different devices, objects, sensors and day-to-day items.

IOT is divided into three main communication models; these models include;

- Device to Device Communication Model: This model describes a scenario whereby two or more devices are connected and communicate between each other, instead of an intermediary application server means.

- Device to Cloud Communication Model: A directly connected device to the internet cloud service is involved in this model, for instance, application service provider for data exchange and message control traffic.
- Device to Gateway Model: In this scenario, the device is connected by an Application Layer Gateway (ALG) service as a gateway to reach the cloud service (Carolyn & Michelle, October 2015)**.**

## 2. Related Works

### 2.1. Fundamental Concept

IoT is described as a system whereby computing devices, objects, digital and mechanical machines, people or animals with the help of a unique identifier have the ability to share data through a network without human to computer or human to human interaction (Rouse, 2020). IoT has advanced since it came into light in the late 1990s. Bill Joy at the world economic forum held in Davis (Switzerland) delivered a speech, he suggested it as one of his proposed six webs taxonomy, which was later used to lecture in many academic and technological forums. The sixth of these webs were called device to device and was described as the internet of sensor connected across a mesh network, setting up urban systems for maximum efficiency (Pontin, 2005).

Kevin Ashton in early 2000s laid the foundation for what the world know today as internet of things at Massachusetts Institute of Technology (MIT) Cambridge United State, automatic identification (AutoID) laboratory. In the RFID journal published in 1999, Ashton described a situation whereby things make use of data they gather without any human interference – by this we will be able to trace and count everything, thereby reducing cost, waste and loss (Ashton, 2009).The International Telecommunication Union (ITU) was also one of the major earliest contributing factors to the understanding of IoT. They describe a system whereby all form of things become functional users of the internet for humans (ITU, 2005). (Atzori et al, 2010) in their analysis, identified three main visions of IoT, which were things oriented, internet oriented and semantic oriented.

Through the advancement of technology, the concept of IOT started adjusting from what can be connected to the web to that, which can be accomplished with things connected to the web. (Stephan et al 2008) claimed the role of IoT is to link physical world to its representation in the information system.

In recent definitions and visions of IoT**,** the part things play as producers of data and network as data enabler of new services were emphasized. (Miorandi et al*, 2012*) explained IoT as an extension of the web and internet into the physical realm. They claimed IoT will use the internet as a worldwide platform to allow smart objects and machines communicate, interact, compute and co-ordinate.

Therefore, IoT describes a system where physical items and sensors within or attached to these items, are connected to the internet via wireless or wired connections.

### 2.2. Related Work

(V. Tamilselvi, 2020) designed a health monitoring system comprising of SpO2, Eye blink, Heartbeat, and Temperature sensors and Arduino-UNO to monitor a patient's vitals. The designed system was bulky and no patient's performance measures were displayed. An IoT health monitoring system was introduced by (AD Acharya, 2020). The proposed system hardware components include Blood Pressure sensor, temperature sensor, ECG sensor, and a mini super computer Raspberry pi. In the proposed system, the data attained from the sensors are processed by the raspberry pi and transmitted to the cloud. The proposed system lacked a data visualization interface to display results. (Alam, 2019) developed an IoT health based monitoring system comprising of Arduino Uno, LM35 temperature sensor, ESP8266-01 Wi-Fi module, pulse sensor and LCD display to monitor the temperature and heartbeat of a patient and display the readings in the internet using ThingSpeak. The major drawback of the developed system is the weight and size which restrict the patient under observation from moving around. (Md. et al, 2015) developed a continuous heart rate and temperature monitoring system using Arduino Uno and android device with an objective to measure the heart rate and body temperature of a patient continuously and then display the result in an android device. The authors used TCRT500 as the heart beat sensor, which senses the change in blood volume of a finger artery while the heart is pumping the blood, LM35 as the temperature sensor for measuring the temperature and HC-05 Bluetooth module which communicate with the Arduino via a serial connection and transfer data to the connected Android device. However, the system was very bulky and complex, the temperature sensor LM35 which is an analog sensor is a good sensor but not reliable. (Kumar et al, 2017) proposed a heath monitoring system which monitors the heart rate and temperature of a patient. However, the system is not fully reliable because each of these sensors considered are made to get reading individually, thereby making the system bulky. (Kale & Mane, 2017) proposed a low power consumption, less complex and portable band for healthcare monitoring system using Arduino Uno and ESP8266 Wi-Fi module. Nonetheless, the system has no alarm system to alert the doctor of emergency.

In this study, we introduced a portable wearable device that can be worn by a patient on his/her wrist. To eliminate the problem of being bulky and the patient being static, we utilized a simple and wearable microcontroller (NodeMCU) which has a Wi-Fi module (ESP8266) embedded in it for transmission to the cloud. We applied a digital temperature sensor (DS18B20) for continuous and accurate temperature readings. A two stage high pass and low pass (Hp-Lp) filter was utilized to eliminate DC offset and noise in the pulse rate sensor readings and a scaling factor (10.41909) was introduced to the digital output of the pulse rate sensor to scale the reading gotten to a readable and easily interpreted data. Also, we made use of ThingSpeak to display the reading gotten by the wearable device in the cloud.

## 3.  System Architecture

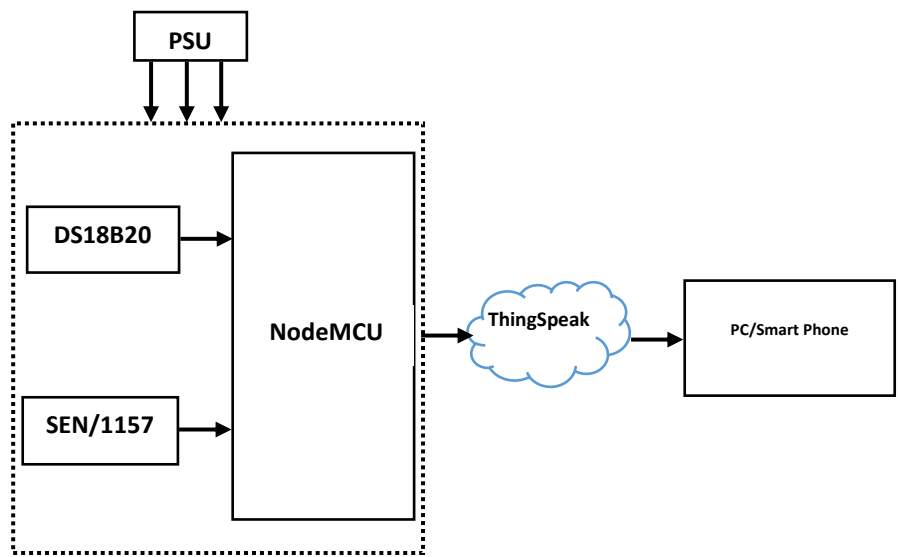

Figure 1: System Architecture

The remote health monitoring system was designed for both the sick and healthy patients. The system utilized IoT to acquire measurement of temperature and pulse rate from the wearable device.

As portrayed in figure 1, the engineering of the proposed remote health monitoring system consists of 3 stages. They include the gathering stage, processing stage, sending stage. During the gathering stage, a wrist band consisting of a temperature and pulse rate sensor was developed to gather information required from patient body vitals. The proposed wrist band is made portable to eliminate the problem of weight as been portrayed by other conventional systems using portable, less dense and cost efficient components. The data gotten from patient's vitals are collected in digital form from the temperature sensor since it's a digital sensor and analog from the pulse rate sensor. In the processing stage, the gathered data or collected data is processed and converted to readable information.

Finally, in the sending stage, the processed data was conveyed through a Wi-Fi module embedded with the microcontroller to the cloud using www.thingspeak.com, so the doctor can access the information at any time and make use of it as long as he has the required login details assigned to the patient.

### 3.1. Description of the Hardware Components

1) Temperature Sensor (DS18B20)

The DS18B20 temperature sensor, a digital sensor temperature measurements ranges from 9bits to 12bits Celsius. It has an alarm system and a nonvolatile user-programmable upper and lower trigger points in it. The DS18B20 tends to communicate through a 1-Wire bus that will require one data line and ground to communicate with the central microprocessor. It operates on a temperature scale ranging from -55°C to +125°C and has an accuracy of ±0.5°C over the range of -10°C to +85°C. This sensor operates on a concept called parasite power i.e. get power directly from the data line, thereby eliminating the need for an external power supply (DFROBOT, 2017).

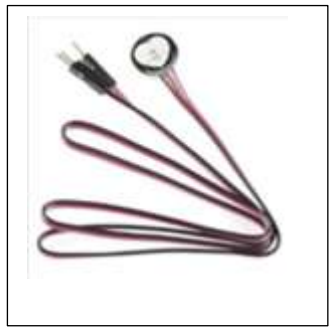

Figure 2: DS18B20 (DFROBOT, 2017)

2) Pulse Rate Sensor (SEN/11574)

This sensor is manufactured to yield a corresponding digital output of body heartbeat when a finger is placed on it. It operate at a voltage of +5V D regulated, and current of 100mA. It also consist of an output data level of 5V TTL level. The pulse beat detection is indicated by a LED and an output high pulse, also has a light source of 660nm super red LED.
SPECIFICATION
Indicator LED
Current: 4mA
Operating voltage: 3.3V – 5V (S. Electronics, 2003)

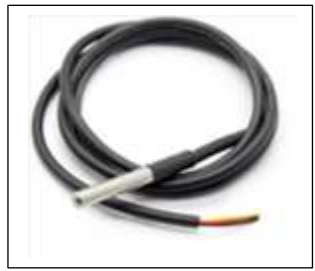

Figure 3: SEN/11574 (S. Electronics, 2003)

3) Microcontroller (NODEMCU ESP8266 ESP-12E)

NodeMCU is an open source for IOT application. It comprises of a firmware that runs on a ESP8266 Wi-Fi SOC from Espressif Systems, with a hardware that is based on the ESP-12 module. The term "NodeMCU" by default relates to the firmware instead of the DevKit. Lua scripting language, that is based on the eLua project, which is built on the Espressif Non-OS SDK for ESP8266 is used for the firmware. It makes use of a lot of open source projects, examples include spiffs and lua-cjson (Einstronic, 2017).

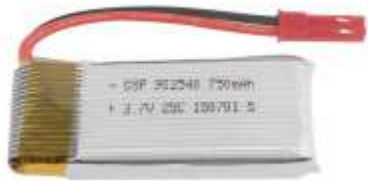

Figure 4: Microcontroller (NodeMCU) (Einstronic, 2017)

4) Lithium Polymer (LiPo) Battery

Lithium Polymer battery is a rechargeable light weight battery that have low chance of suffering from leakage of electrolyte when compared to other Lithium batteries. Lithium Polymer (LiPo) batteries are generally flexible and robust (Salt, 2008-2018).

To implement the concept of a portable wearable device, a LiPo battery was adopted. LiPo Battery meets the power requirement needed to power the portable wearable device. It has a nominal voltage of 3.7V DC and a capacity of 750mAh. When fully charged can last up to 6 hours. It net weight is 14g and size of 50*30*5mm, which makes it portable and weighless when utilized in the implementation process.

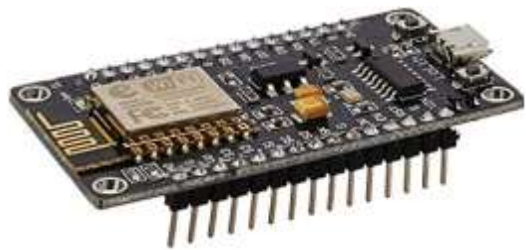

Figure 5: Lipo Battery (Salt, 2008-2018)

## 3.2. Implementation

The implementation procedure was divided into two parts: the hardware part and software part:

*Hardware Implementation Process*

The hardware implementation process consists of interfacing the digital sensor (DS18B20) and analog sensor (SEN/11574) with the microcontroller (NodeMCU). The temperature sensor (DS18B20) has three pins, the second pin which is the output is connected to the digital pin of the microcontroller (NodeMCU), the first pin which is the ground GND is connected to the ground GND of the microcontroller (NodeMCU) and the third pin which is the power VCC to the power of the microcontroller (NodeMCU). The digital pin of the microcontroller is 10 bits. The supply voltage from the microcontroller to the sensor is 3.3V and the VCC and output are linked with 4.7K resistor. The Pulse Sensor (SEN/11574) is connected to microcontroller as follows, the signal pin to ADC0, the VCC pin to 3.3V, the GND pin to the GND of the microcontroller. The system is then powered by the use of a rechargeable 3.7V 750mAh Lithium Polymer (LiPo) Battery.

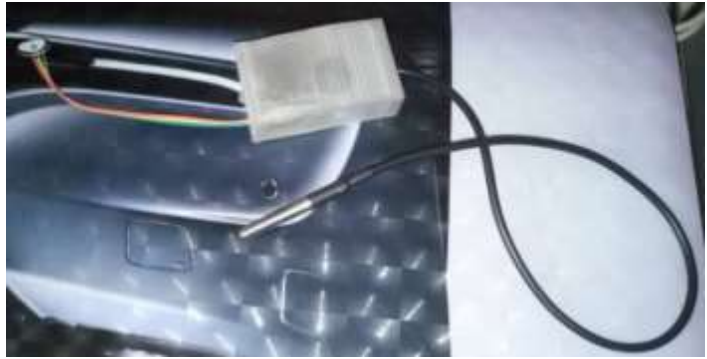

Figure 6: Experimental Setup

*Software Implementation Process*

The software implementation process consists of the coding aspect which was written on Arduino IDE (Integrated Development Environment), then complied to check for errors and after which, uploaded to the device through a USB 2.0 cable. The programming language used in the implementation of the device is C programming language. C programming language was adopted because it is user friendly and easiness to use while coding a microcontroller. The code reads raw values from the temperature and pulse rate sensor of the proposed device, the readings from the pulse rate sensor is processed using a high and low pass filters to measure and analyze the accurate pulse rate data. In order to accomplish this, the signal measured from the SEN/11574 in form of an electrical signal which consist of a DC signal that correspond to the blood volume and tissue and AC component synchronous to the pulse rate is filtered through a two stage HL-LP circuit to attain the AC signal which is of significant importance. This filtering is necessary to eliminate the presence of noise and DC offset that causes deviation in the peak during the measurement to attain a desired frequency spectrum of approximately 1Hz – 3Hz for both an idle and active pulse rate (CORP, 2012). The result of the HP-LP circuit is then converted to digital pulses utilizing a comparator circuit or ADC.

$$BPM \text{ (Beats per minute)} = 60*F$$

Because of the high range of deviation in the readings, a scaling factor is required to scale the yield to a typical meaningful output range**.** The scaling factor 10.41909 for the pulse rate readings was gotten by comparing the readings gotten during the course of testing with another reading gotten through the use of a Samsung S7 Edge. This Samsung S7 Edge has a pulse rate sensor embedded in it. Ten individual readings were gotten from the two different sensor simultaneously. The average of this individual readings were calculated and then interpolated. The average reading of the designed system pulse rate sensor was 802.27, while the average reading of the Samsung S7 Edge was 77.

Therefore, Scaling Factor = $^{802.27}/_{77}$ = 10.41909

This digital pulses are feed to the microcontroller and used to calculate the heartbeat rate;

$$BPM \text{ (Beats per minute)} = 60*F/(10.41909)$$

The calculation and analysis of the sensors play a major role in the software design because system failures such as data loss and wrong data measurement can occur if not implemented properly.

When the cloud connection is lost, there is no data loss because the wearable device saves the data in the microcontroller memory pending when a connection is established. The online platform used in the viewing of the readings gotten from the system is www.thingspeak.com.

ThingSpeak is an online platform that provides a good tool for IoT based projects. ThingSpeak website, allows the monitoring and controlling of the data produced by our system through the Internet, using the webpages and channels provided by the platform. ThingSpeak acquires the data produced by the sensors or actuators, analyzes and display them and also acts by triggering a reaction. ThingSpeak has previously been utilized in different IoT projects implementation (Jain, 2019). In order to gain access to the website, a new account was registered and channel consisting of two fields created, one field for displaying the temperature readings in degree Celsius ($^{o}$C) and the other field for displaying the pulse rate readings in beat per minutes (bpm). For this channel, a unique API key was given, which in turn was put in the code using Arduino IDE and uploaded to the microcontroller board NodeMCU.

To deal with the issues of security, ThingSpeak provides a login forum as shown in Figure 7 below. The username and password are known only to the patient and doctor or specialist administering medical care to such patient for security purpose and this detail enable them to view data (readings) of patients. This makes it less vulnerable to hacking and leaking of a patient's medical information.

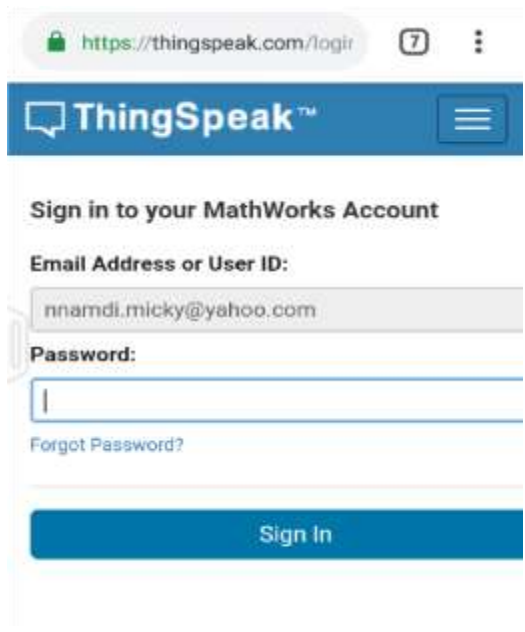

Figure 7: ThingSpeak login forum

Figure 8 above shows the flowchart that the system follows. The system repeats the process of acquiring new sets of patient's body vitals after every 30 minutes. The set of new data is uploaded to the cloud automatically after being gathered and processed, it is also saved in the microcontroller memory.

### 3.3. Mode of Operation

*I. Principle of Temperature Sensor (DS18B20) Operation*

The temperature sensors basic principle of operation is its voltage across the diode terminals. The voltage across the terminal is indirectly proportional to the temperature i.e. when the voltage increases, the temperature also does, followed by a drop in the voltage between the emitter in the diode and the transistor terminal. The main functionality of this temperature sensor (DS18B20) is its ability to produce digital readings. The default resolution at power-up is 12-bit. The temperature measurement and A-to-D conversion is done by a convert-T command. After this conversion, the corresponding readings are stored in the 2-byte register of the sensor after which, the sensor returns to its idle state

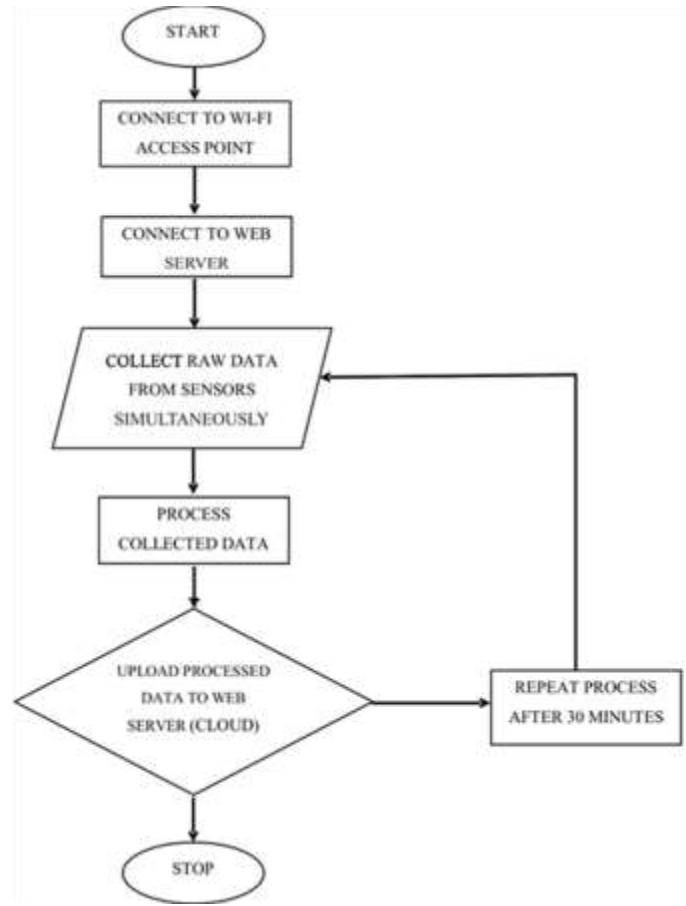

Fig 8 Flow diagram

## II.  Principle of Pulse Rate Sensor (SEN/11574) Operation

The pulse rate sensor based its principle on photoplethysmography. It measures the change in volume of blood passing through any body organ which tend to cause an effect within the light tendency through that body organ (avascular region). SEN/11574 consist of a light emitting diode and a photodiode (light detecting resistor). The heartbeat pulses are responsible for the variation that occurs within the stream of blood to various area in the body. When tissue is illuminated with the light emitted by the led, it reflects (a finger tissue). A portion of the light is consumed by the blood and transmitted or the reflected light is gotten by the light detector. The amount of light retained relies upon the blood volume in that particular tissue. The yield of the detector comprises of an electric signal which consist of DC offset, high frequency noise and AC components. This electrical signal is passed through a two stage Hp-Lp filter to eliminate the high frequency noise and DC offset. The resultant desired frequency is converted to digital pulses using a ADC and then scaled down using the digit '10.41909' which was attained through the comparison of the readings gotten simulateneously from a Samsung Edge and the introduced system pulse rate sensor (SEN/11574) to a readable range.

Mathematically,

BPM (Beats per minute) = 60*F/(10.41909)

### 3.4. System Methodology

When the system is being switched on through the switch, the battery supplies the microcontroller (NodeMCU) with a supply voltage of 3.3V-3.7V to power it. In turn, the microcontroller (NodeMCU) powers the digital temperature sensor (DS18B20) and the analog pulse rate sensor (SEN/11575) by connecting their respective voltage common collector (VCC) and ground (GND) pins to the microcontroller. The data generated by this device is not presented to the doctor directly but rely on machine to machine (M2M) communication protocol to get the data to the point where it is processed and used in relevant ways. The DSI8B20 and SEN11574 detect or sense the patient temperature and heart rate readings simultaneously, one in an analogous way and the other in a digital way and sends the raw data gotten from the pulse rate sensor to an analog to digital converter (ADC) embedded in the microcontroller (NodeMCU), this ADC does the conversion of the collected data which is in voltage form into digital numbers equivalent. While the readings gotten from the DS18B20 need not be converted since its readings are already gotten in digital form. Inside the NodeMCU is an integrated development environment (IDE) where programs or codes are written. This code help process the digital signal from the ADC and digital signal gotten directly from the digital sensor to machine language since the computer only understands 1 and 0. Also embedded inside the NodeMCU is a WIFI module which has transmission control protocol (TCP), it finds and connect the Wi-Fi Access Point and this helps to transmit the processed information to the cloud (THINGSPEAK) and then the doctor can access it from anywhere with the necessary login information.

## 4. Testing and Results

The readings from different persons were taken during testing, and several readings were recorded.

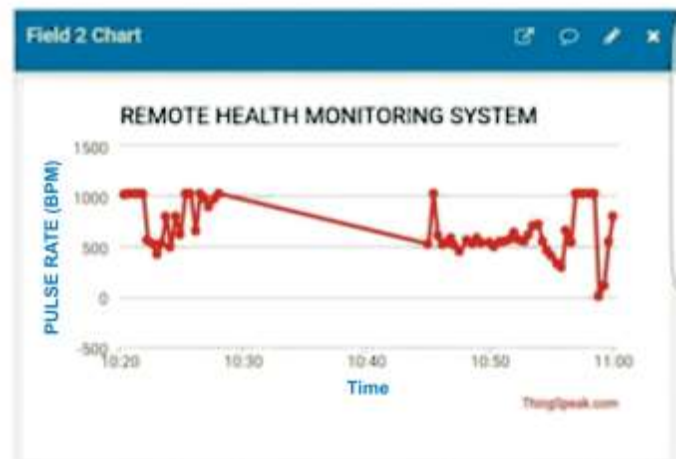

Figure 8: Graphical Representation of Patient's Pulse Rate

Figure 9 above is an analytical graphical representation of the pulse rate of a patient taken around 10:45 – 11:00, July 2, 2018 which is presented in Deci-Beats per Minute and can be further converted to Beats per Minute by dividing it by (10.41909). Scaling Factor = $^{802.27}/_{77}$ = 10.41909

This scaling factor 10.41909 was gotten from the comparison with a Samsung S7 pulse rate readings gotten simulteneouly with the designed system pulse rate sensor (SEN/11574). The scaling factor was inputted into the program code to scale it down automatically giving an accurate reading in beats per minute (Bpm).

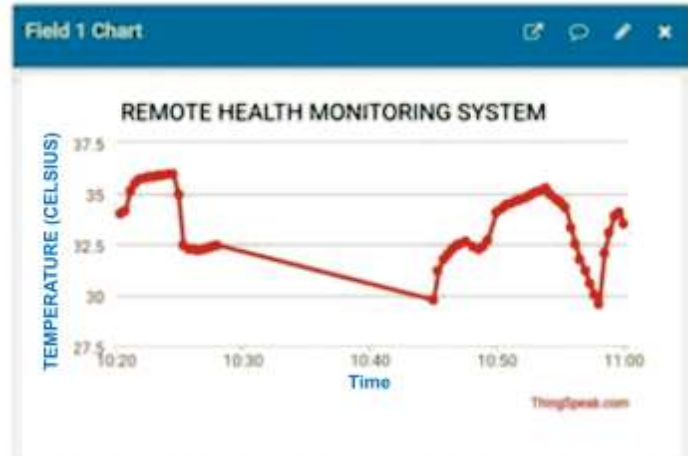

Figure 9: Graphical Representation of Patient's Body Temperature

Figure 10 above shows the graphical representation of the body temperature presented in degree Celsius of the same patient also taken around the same interval as the pulse rate readings. Figure 11 above shows the pulse rate of same patient taken 04 July, 2018 in the space of 30 minutes, the readings gotten this time are actually represented in actual beats per minutes (BPM) and the doctor need not divide it.

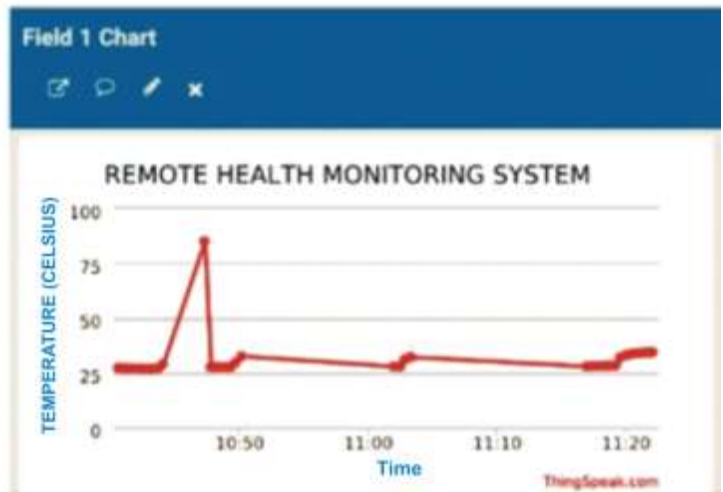

Figure 11 shows the temperature readings of same patient within the same date and time frame.

## 5. Conclusion

The main objective of the project was successfully achieved. The two individual sensors (temperature sensor - DS18B20 and Pulse rate sensor - SEN/11574) gave out the required results simultaneously with a high success precision. An online platform "www.thingspeak.com" was utilized to access the readings on the internet by a doctor or even the patient, if with the necessary logging details. The constructed device helps in bridging the gap between a doctor and his patients. The use of NodeMCU, a microcontroller with Esp 8266 Ep-12E embedded in it made it easy to reduce the size since there was no need for an external Wi-Fi module to transfer the readings gotten to the cloud (www.thingspeak.com). The designed system consists of a wearable portable band for health monitoring with low power consumption. Some more measures which are very significant to determine a patient's condition like heart diseases, blood pressure, respiration monitoring, diabetes etc. can be looked into as future work.